\documentclass[12pt,preprint,showkeys,indentfirst,bm,superscriptaddress]{revtex4}
\usepackage{epsfig}
\usepackage{epsf}
\usepackage{bm}
\topmargin=1.0cm \Roman{table}

\newcommand{\al}{&\!\!\!\!~}
\newcommand{\be}{\begin{equation}}
\newcommand{\ee}{\end{equation}}
\newcommand{\ba}{\begin{eqnarray}}
\newcommand{\ea}{\end{eqnarray}}
\newcommand{\noo}{\nonumber}

\begin{document}


\title{The $\pi\pi$ Phase Shifts from $\psi'\to J/\psi\pi^+\pi^-$ Decays}

\author{GUO Feng-Kun$^{1,6}$}
\email{guofk@mail.ihep.ac.cn}
\author{SHEN Peng-Nian$^{1,2,4,5}$}
\author{JIANG Huan-Qing$^{3,1,4,5}$\\
{\small $^1$Institute of High Energy Physics, Chinese Academy of Sciences,Beijing 100049, China}\\
{\small $^2$CCAST(World Lab.), P.O.Box 8730, Beijing 100080, China}\\
{\small $^3$South-west Normal University, Chongqing 400715, China}\\
{\small $^4$Institute of Theoretical Physics, Chinese Academy of Sciences, Beijing 100080, China}\\
{\small $^5$Center of Theoretical Nuclear Physics, National
Laboratory of Heavy Ion Accelerator,}\\
{\small Lanzhou 730000, China}\\
{\small $^6$Graduate School of the Chinese Academy of Sciences,
Beijing 100049, China}}


\vspace{1cm}

\begin{abstract}
 The $\psi'\to J/\psi\pi^+\pi^-$ decay process provides a new way to extract
 the $\pi\pi$ $S$ wave phase shifts up to $0.59GeV$. In this paper we derive the
formulae for extracting  the $\pi\pi$ $S$ wave phase shifts from
the invariant mass spectrum of $\pi\pi$ in the $\psi'\to
J/\psi\pi^+\pi^-$ decay.

\end{abstract}


\maketitle

Meson-meson scattering is an important process for understanding
the fundamental hadron-hadron interactions. $\pi\pi$ interaction,
as one of the simplest process, has substantially been studied
experimentally and theoretically over many years. Up to now, the
theoretical study of the $\pi\pi$ interaction has been made using
perturbative \cite{chpt} and non-perturbative chiral effective
theories \cite{tr,chut,xz}, boson exchange models\cite{zou}, Regge
analysis \cite{reg} and precise dispersion relation analysis
\cite{dpr}. The experimental investigations on this line were
commonly carried out in terms of the reactions dominated by one
pion exchange (for a review, see \cite{mms}, for example), and the
$\pi\pi$ phase shifts were extracted from the analysis of the
reactions $\pi N\to \pi\pi N$ (e.g. $\pi^- p\to n\pi^+\pi^-$
\cite{hj73,gh74,em74,fp77}
 and $\pi^- p\to
n\pi^0\pi^0$ \cite{as03}), $\pi^+ p\to \Delta^{++}\pi^+\pi^-$ and
$\pi^+ p\to p\pi^+\pi^0\pi^0$ \cite{bc81}. The off-shell
extrapolation of the $\pi\pi$ interaction and the reaction
mechanism cause  uncertainties for the determination of the
$\pi\pi$ phase shifts.

Many years ago, it was realized that when particles are produced
in a reaction, some of these particles often interact strongly
with each other before going outside the interaction range
\cite{wa52}, and this, so called final state interaction (FSI),
makes the analysis complicated. But, FSI does provide important
information about the reaction mechanism and the interaction among
the outgoing particles.

In this paper we discuss the  $S$-wave isoscalar-channel $\pi\pi$
scattering in the $\psi'\to J/\psi\pi^+\pi^-$ decay process and
provide a formalism for extracting the $\pi\pi$ $S$ wave phase
shifts from the invariant mass spectrum of the  $\pi\pi$ in the
$\psi'\to J/\psi\pi^+\pi^-$ decay.

For the decay process $\psi'\to J/\psi\pi^+\pi^-$, only $I=0$
channel contributes to the final $\pi\pi$ amplitude because the
isospins of $\psi'$ and $J/\psi$ are zero. This isospin selection
makes $\pi\pi$ phase shift analysis simpler and clearer. In the
analysis, the phase space limits the kinematic region of the
$\pi\pi$ invariant mass in a region below $0.59GeV$, and
Bose-statistics limits the angular-momentum between two pions to
be 0 or 2, i.e., $S$ wave or $D$ wave. Because the $D$ wave
component in the $\pi\pi$ final state is smaller compared to the
$S$ wave component\cite{bes00}, which is consistent with simple
analysis, for simplicity, we ignore the $D$ wave component and
consider the $S$ wave $\pi\pi$ FSI only in the analysis. As
mentioned in Refs.~\cite{bc75,mu97}, the Lagrangian for the
$\psi'\to J/\psi\pi^+\pi^-$ decay in Fig.1(a) can be written in a
form of contact term
\ba%
    g\psi'_{\mu} \psi^{\mu}\partial_{\nu}\phi\partial^{\nu}\phi'
\ea%
Due to FSI, one has to consider an additional diagram Fig.1(b) for
the $\psi'\to J/\psi\pi^+\pi^-$ decay. Let ${\bm \beta_i}$ and
$m_i$ denote the velocity and the mass of particle $i$ in the
final state, respectively. It is easy to derive a relation between
${\bm \beta_i}$ and $m_i$
\ba%
{\bm \beta_i}^2\leq 1-\frac{m_i^2}{(M-\sum_{j\neq i} m_j)^2},
\ea%
where $M$ is the mass of the initial particle, $\psi'$ in this
case. This inequality shows that the possible maximal velocity of
$J/\psi$ is much smaller than the corresponding pions'. Therefore,
we can neglect the FSI between $J/\psi$ and pions.

\begin{figure}[htb]
 \includegraphics{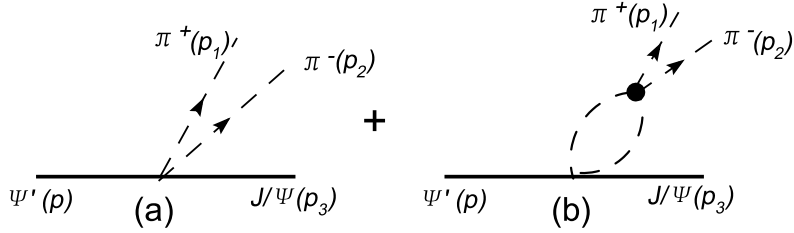}%
 \caption{\label{fig:feyn}Diagrams for the $\psi'\to
 J/\psi\pi^+\pi^-$ decay. (a) represents the contact interaction, and (b) denotes
  the process with $\pi\pi$ FSI included.}
\end{figure}
The decay amplitude for Fig.~\ref{fig:feyn} can be written in
terms of $\pi\pi$ amplitudes obtained by using the factorization
approximation
\be%
\label{eq:t} T\equiv\langle J/\psi\pi^+\pi^-|t|\psi'\rangle =
P (1 + G t^{I=0}_{\pi^+\pi^-, \pi^+\pi^- + \pi^-\pi^+ + \pi^0\pi^0}),%
\ee%
where $P$ is the amplitude of the contact term, $G$ represents the
loop propogator, and $t^{I=0}_{\pi^+\pi^-, \pi^+\pi^- + \pi^-\pi^+
+ \pi^0\pi^0} = \langle \pi^+\pi^-|t|\pi^+\pi^- + \pi^-\pi^+
+\pi^0\pi^0\rangle$ is the amplitude of the process $\pi^+\pi^- +
\pi^-\pi^+ +\pi^0\pi^0 \to \pi^+\pi^-$. If we denote the
four-momenta of $\psi'$, $\pi^+$, $\pi^-$ and $J/\psi$ by $p$,
$p_1$, $p_2$ and $p_3$, respectively, the term
$PGt^{I=0}_{\pi^+\pi^-, \pi^+\pi^- + \pi^-\pi^+ + \pi^0\pi^0}$
actually means an integration%
\be%
\int \frac{d^4q}{(2\pi)^4} P(p,p_3;q)G(p-p_3,q)t(q;p_1,p_2),%
\ee%
where $q$ represents the pion momentum on one of the loop lines.

The amplitude for the $S$ wave contact term reads \cite{bc75,mu97}
\ba%
\label{eq:p} P(s) \al=\al
-\frac{4g}{f^2_{\pi}}\epsilon'^{(\lambda')}\cdot\epsilon^{(\lambda)*}p_1\cdot
p_2 \noo \\
\al=\al -\frac{2g}{f^2_{\pi}}\epsilon'^{(\lambda')}\cdot\epsilon^{(\lambda)*}(s-2m_{\pi}^2),%
\ea%
where $g$ is the coupling constant and $f_{\pi}$ is the pion decay
constant with the experimental value of $92.4MeV$. From
Eq.~(\ref{eq:p}), one sees that $P$ is not $q$-dependent. this
implies the direct production term $P$ can be factorized from the
loop integration.

With the phase convention $|\pi^+\rangle =-|1,1\rangle$, the wave
function for the $I=0$ $\pi\pi$ system can be written as
\be%
|\pi\pi\rangle^{I=0} =
-\frac{1}{\sqrt{6}}(|\pi^+\rangle|\pi^-\rangle +
|\pi^-\rangle|\pi^+\rangle + |\pi^0\rangle|\pi^0\rangle).%
\ee%
Then the amplitude $t^{I=0}_{\pi^+\pi^-, \pi^+\pi^- + \pi^-\pi^+ +
\pi^0\pi^0}$ can be rewritten as
\ba%
\langle \pi^+\pi^-|t|\pi^+\pi^- + \pi^-\pi^+ +\pi^0\pi^0\rangle
\al=\al
\langle \pi^+\pi^-|t|(-\sqrt{6})\pi\pi\rangle^{I=0} \noo \\
\al=\al ^{I=0}\langle (-2/\sqrt{6})\pi\pi|t|(-\sqrt{6})\pi\pi\rangle^{I=0} \noo \\
\al=\al 2t^{I=0}_{\pi\pi,\pi\pi},%
\ea%
where $t^{I=0}_{\pi\pi,\pi\pi}$ is the $I=0$ $\pi\pi$ scattering
amplitude which can be related to the $S$ wave $I=0$ $\pi\pi$
scattering phase shifts if we neglect the $D$ wave contribution.
Note that with the normalization
$^{I=0}\langle\pi\pi|\pi\pi\rangle^{I=0}=1$,
$|\pi^+\rangle|\pi^-\rangle$ should be normalized to 2.

On the other hand, we take the normalization for partial wave
amplitudes in such a way that the unitary relation for the partial
amplitudes satisfies
\ba%
Im T_l(s)=T^{\dagger}_l(s)\rho(s)T_l(s)%
\ea%
with the phase space factor%
\be%
\rho(s)=\frac{p_{cm}}{8\pi
\sqrt{s}}=\frac{1}{16\pi}(1-\frac{4m_{\pi}^2}{s})^{1/2}.%
\ee%
The relation among the partial wave amplitude $T^I_l$ with isospin
$I$ and the phase shift parameters $\delta^I_l$ and $\eta^I_l$
reads
\ba%
T^{I}_l(s)=\frac{1}{2i\rho(s)}(\eta^I_l(s) e^{2i\delta^I_l(s)}-1).%
\ea%
Then, taking on-shell approximation in relating
$t^{I=0}_{\pi\pi,\pi\pi}$ in the decay amplitude to the physical
values of phase shifts, the $S$ wave isoscalar $\pi\pi$ scattering
amplitude $t^{I=0}_{\pi\pi}$ becomes
\ba%
\label{eq:ps}
t^{I=0}_{\pi\pi,\pi\pi}(s)=\frac{1}{2i\rho(s)}(e^{2i\delta^0_0(s)}-1),%
\ea%
where the inelastic coefficient $\eta^0_0(s)$ is taken to be 1 in
this elastic scattering process. From Eq.~(\ref{eq:ps}), one finds
that $t^{I=0}_{\pi\pi,\pi\pi}$ is also independent of $q$, so that
it can be factorized out from the loop integration. The $\pi\pi$
loop integration
\begin{equation}
G(s)=i\int\frac{dq^4}{(2\pi)^4}\frac{1}{q^2-m_{\pi}^2+i \varepsilon} \frac{1}{%
(p_1+p_2-q)^2-m_{\pi}^2+i\varepsilon},  \label{eq:2loop}
\end{equation}
where $s=(p_1+p_2)^2$ is the squared four-momentum of the dipion
system, can be calculated by using a three-momentum cut-off
parameter $q_{max}$, and the analytic formula in the frame of the
center of mass of dipion can be derived as
\begin{equation}
\label{eq:g} G(s)=\frac{1}{8\pi^2}\{\sigma(s)
\arctan{\frac{1}{\lambda\sigma(s)}} -
\ln[\frac{q_{max}}{m_{\pi}}(1+\lambda)]\},
\end{equation}
where $\sigma(s)=\sqrt{\frac{4m_{\pi}^2}{s}-1}$,
$\lambda=\sqrt{1+\frac{m^2_{\pi}}{q^2_{max}}}$, and $q_{max}$
takes a reasonable value around $1GeV$ \cite{mg84,oo97}.

Finally, we rewrite the decay amplitude as
\ba%
\label{eq:tp} T \al\equiv\al \epsilon'^{(\lambda')}\cdot\epsilon^{(\lambda)*}T_R(s) \noo \\
\al=\al
-\frac{2g}{f^2_{\pi}}\epsilon'^{(\lambda')}\cdot\epsilon^{(\lambda)*}(s-2m_{\pi}^2)
(1 + 2G(s) t^{I=0}_{\pi\pi,\pi\pi}).%
\ea%
From Eq.~(\ref{eq:tp}), one finds that the angular dependence in
the total decay amplitude comes from
$\epsilon'^{(\lambda')}\cdot\epsilon^{(\lambda)*}$ only.

The differential decay width is given \cite{pdg04} by
\be%
d\Gamma=\frac{1}{(2\pi)^5}\frac{1}{16M^2}\sum_{\lambda}\overline{\sum_{\lambda'}}|T|^2|{\bf
p_1^*}||{\bf
p_3}|dm_{\pi\pi}d\Omega^*_1d\Omega_3,%
\ee%
where $M$ is the mass of $\psi'$, $m_{\pi\pi}=\sqrt{s}$ is the
invariant mass of $\pi^+\pi^-$,
$\sum_{\lambda}\overline{\sum}_{\lambda'}$ describes the average
over initial states and the sum over final states, $(|{\bf
p_1^*}|, \Omega^*_1)$ is the momentum of $\pi^+$ in the rest frame
of the dipion, and $(|{\bf p_3}|, \Omega_3)$ is the momentum of
$J/\psi$ in the rest frame of $\psi'$. $|{\bf p_1^*}|$ and $|{\bf
p_3}|$ are expressed, respectively, by
\be%
|{\bf p_1^*}| = (\frac{s}{4}-m_{\pi}^2)^{1/2}, \ee and \be
|{\bf p_3}| = \frac{1}{2M}[(M^2-(m_{\pi\pi}+m_3)^2)(M^2-(m_{\pi\pi}-m_3)^2)]^{1/2},%
\ee%
where $m_3$ is the mass of $J/\psi$. Taking into account the fact
that $\psi'$ produced in the $e^+e^-$ collision experiments is
transversely polarized with respect to the beam direction, the
average over the initial state should be performed with respect to
only two directions. With
\ba%
\sum_{\lambda}\overline{\sum_{\lambda'}}|\epsilon'^{(\lambda')}\cdot\epsilon^{(\lambda)*}|^2
\al=\al \frac{1}{2}\sum_{\lambda=0,\pm1}\sum_{\lambda'=\pm1}|\epsilon'^{(\lambda')}\cdot\epsilon^{(\lambda)*}|^2 \noo \\
\al=\al 1+\frac{{\bf p_3}^2}{2m_3^2}(1-\cos^2\theta_3),%
\ea%
the differential decay width with respect to the $\pi\pi$
invariant mass can finally be written as
\ba%
\frac{d\Gamma}{dm_{\pi\pi}} \al=\al
\frac{1}{32\pi^3M^2}|T_R(s)|^2|{\bf p_1^*}||{\bf
p_3}|(1+\frac{{\bf
p_3}^2}{3m_3^2}) \noo \\
\al=\al \frac{4g^2}{32\pi^3M^2f^2_{\pi}}(s-2m_{\pi}^2)^2|{\bf
p_1^*}||{\bf p_3}|(1+\frac{{\bf
p_3}^2}{3m_3^2})|1+\frac{G(s)}{i\rho(s)}(e^{2i\delta^0_0(s)}-1)|^2,%
\ea%
where $s=m_{\pi\pi}^2$. By fitting experimental $\pi^+\pi^-$
invariant mass spectrum of the $\psi'\to J/\psi\pi^+\pi^-$decay
bin by bin, one can extract the $S$ wave isoscalar $\pi\pi$ phase
shifts.

We thank Prof. Zhu Yong Sheng and Drs. Li Gang and Wu Ning for
their suggestions and discussions. This work is partial supported
under the NSFC grant Nos. 90103020, 10475089, 10435080 and the CAS
Knowledge Innovation Key-Project grant No. KJCX2SWN02.

\end{document}